\documentclass[10pt]{IEEEtran}
\IEEEoverridecommandlockouts
\usepackage{pdfx}
\usepackage{cite}
\usepackage{amsmath,amssymb,amsfonts}
\usepackage{algorithmic}
\usepackage{tabularx}
\usepackage{adjustbox}
\usepackage{multirow}
\usepackage{makecell}
\usepackage{graphicx}
\usepackage{textcomp}
\usepackage{comment}
\usepackage{xcolor}
\usepackage{cleveref}
\usepackage{braket}
\usepackage{enumitem}
\usepackage{tikz}
\def\BibTeX{{\rm B\kern-.05em{\sc i\kern-.025em b}\kern-.08em
    T\kern-.1667em\lower.7ex\hbox{E}\kern-.125emX}}

\usepackage[compact]{titlesec}

\titlespacing\section{0pt}{0.15\baselineskip}{0.1\baselineskip}
\titlespacing\subsection{0pt}{0.1\baselineskip}{0.05\baselineskip}
\titlespacing\subsubsection{0pt}{0.075\baselineskip}{0.05\baselineskip}

\newcommand*\circled[1]{\tikz[baseline=(char.base)]{
            \node[shape=circle,draw,inner sep=0.8pt, minimum size=2pt] (char) {#1};}}
\newcommand{\rpoint}[1]{\circled{{\fontfamily{pcr}\selectfont\footnotesize{#1}}}}

\usepackage{fancyhdr,lipsum}
\fancypagestyle{firstpage}{
  \fancyhf{}
  \fancyhead[C]{To appear at the 2025 International Joint Conference on Neural Networks (IJCNN), Rome, Italy, July 2025.}
  \fancyfoot[C]{\thepage}
}

\begin{document}

\title{Noisy HQNNs: A Comprehensive Analysis of Noise Robustness in Hybrid Quantum Neural Networks
}

\author{\IEEEauthorblockN{Tasnim Ahmed\IEEEauthorrefmark{1}\IEEEauthorrefmark{2}, Alberto Marchisio\IEEEauthorrefmark{1}\IEEEauthorrefmark{2}},
Muhammad Kashif \IEEEauthorrefmark{1}\IEEEauthorrefmark{2},
Muhammad Shafique\IEEEauthorrefmark{1}\IEEEauthorrefmark{2}

\IEEEauthorblockA{\IEEEauthorrefmark{1}  eBrain Lab, Division of Engineering, New York University Abu Dhabi, PO Box 129188, Abu Dhabi, UAE}\\
\IEEEauthorblockA{\IEEEauthorrefmark{2} \normalsize Center for Quantum and Topological Systems, NYUAD Research
Institute, New York University Abu Dhabi, UAE}

Emails: \{tasnim.ahmed, alberto.marchisio, muhammadkashif, muhammad.shafique\}@nyu.edu

\vspace{-20pt}
}

\maketitle
\thispagestyle{firstpage}

\begin{abstract}
Hybrid Quantum Neural Networks (HQNNs) offer promising potential of quantum computing while retaining the flexibility of classical deep learning. However, the limitations of Noisy Intermediate-Scale Quantum (NISQ) devices introduce significant challenges in achieving ideal performance due to noise interference, such as decoherence, gate errors, and readout errors. This paper presents an extensive comparative analysis of two HQNN algorithms, Quantum Convolutional Neural Network (QCNN) and Quanvolutional Neural Network (QuanNN), assessing their noise resilience across diverse image classification tasks. We systematically inject noise into variational quantum circuits using five quantum noise channels: Phase Flip, Bit Flip, Phase Damping, Amplitude Damping, and Depolarizing Noise. By varying noise probabilities from 0.1 to 1.0, we evaluate the correlation between noise robustness and model behavior across different noise levels.

Our findings demonstrate that different noise types and levels significantly influence HQNN performance. The QuanNN shows robust performance across most noise channels for low noise levels (0.1 - 0.4), but succumbs to diverse effects of depolarizing and amplitude damping noise at probabilities between (0.5 - 1.0). However, the QuanNN exhibits robustness to bit flip noise at high probabilities (0.9 - 1.0). On the other hand, the QCNN tends to benefit from the noise injection by outperforming noise-free models for bit flip, phase flip, and phase damping at high noise probabilities. However, for other noise types, the QCNN shows gradual performance degradation as noise increases. These insights aim to guide future research in error mitigation strategies to enhance HQNN models in the NISQ era.

\end{abstract}

\begin{IEEEkeywords}
 Quantum Machine Learning, Hybrid Quantum-Classical Neural Networks, Quantum Convolutional Neural Networks, Quanvolutional Neural Networks, Noise Robustness
\end{IEEEkeywords}

\section{Introduction}

Quantum computing (QC) is the next generation computing paradigm that utilizes quantum mechanical principles to enhance problem-solving capabilities in multiple areas~\cite{google_qc, ibm_qc, qc_adv1}. One field where QC has demonstrated a significant advantage over classical computers is Quantum Machine Learning (QML)~\cite{Schuld_QML_HS, Markidis_ProgrammingQNNs,  Massoli_LeapQCQNN, Huang_2021a}. This emerging field combines machine learning (ML) and QC to address challenges associated with high-dimensional data processing and analysis of complex tasks~\cite{qml_intro, qml}. QML seeks to exploit the unique properties of quantum systems to develop advanced algorithms, neural network architectures, and training methodologies\cite{kashif:2023impact}. The potential applications of QML span various domains, including finance, health care, and computer vision~\cite{innan2024fedqnn, innan2024fraud, innan2024qfnn, zaman2024po, siddiqui2024quantum, innan2024lep, innan2024fl, el2024quantum, dutta2024qadqn, Schetakis_ReviewQML, qml_health}. 

However, the current state of quantum technology presents significant challenges. We are currently in the Noisy-Intermediate Scale Quantum (NISQ) era, characterized by quantum processors with a limited number of qubits that are highly susceptible to errors, such as decoherence, gate errors, and readout errors~\cite{Preskill, yu2023efficientseparatequantificationstate}. These limitations hinder the practical realization of standalone QML algorithms~\cite{qml_challenge, zaman2023survey}. 

To address these challenges, hybrid quantum-classical machine learning (HQML) models have been developed to operate efficiently within the compatibility constraints of NISQ devices~\cite{pennylane_diff, marchisio2024cutting}. Different HQML models have been proposed, including quantum support vector machines, quantum Boltzmann machines, quantum auto-encoders, and quantum reservoir computing~\cite{Regression:2022, Havl_ek_2019, Kubler:2021, reservoir_qc, Amin_2018, generative:2018}. Among these, hybrid quantum neural networks (HQNNs) have garnered significant interest due to parameterized quantum circuits (PQCs), which offer high expressiveness and can potentially reduce the number of trainable parameters compared to classical neural networks when implemented with appropriate encoding techniques~\cite{Kashif_HQNN:2022, Abbas:2021, Ghasemian2022, farhi2018classification, kashif2024computational, Kashif_NRQNN, maouaki2024designing}. The learning process of HQNNs typically follows these steps: (1) pre-processing classical data input, (2) encoding classical data into qubit states, (3) training the PQCs on encoded data, (4) measuring the final states of the qubits, and (5) post-processing the measurement results of PQCs classically~\cite{Kashif_HQNN:2022, kashif:2021_DSE}. 

Despite their promise, HQNNs remain highly sensitive to quantum noise, which can degrade their accuracy and reliability. Extensive research work has explored the circuit design and training methodologies~\cite{zaman2024comparative, HurQCNN:2022, zaman2024studying, Kashif_2024_resqnets, el2024advqunn, maouaki2025qfal}. However, most studies primarily evaluate the behavior and performance of algorithms under ideal noise-free conditions, overlooking the impact of quantum noise on their practical development into NISQ devices. A comprehensive understanding of the interplay between noise and HQNN circuit architecture in NISQ devices remains a highly needed step towards mitigating these noise effects and realizing the full potential of QML. 

While previous work has explored noise impacts on specific quantum algorithms, specific quantum noise sources, or simplified HQNNs~\cite{kashif2024investigating, qnn_motten, reservoir_qc, ahmed2025quantumneuralnetworkscomparative, kashif2024:HQNET, el2024robqunns}, a thorough analysis of how diverse noise channels affect more complex HQNN architectures in multi-classification tasks is still needed.

This paper analyzes the performance and noise resilience of two widely studied HQNN models for image classification: Quanvolutional Neural Networks (QuanNN)\cite{QuanNN:2019,kashif:2024resqunns} and Quantum Convolutional Neural Networks (QCNN)\cite{QCNN:2019}. Both models draw inspiration from classical Convolutional Neural Networks (CNNs), but employ quantum circuits differently. The QuanNN uses a single quantum circuit as a filter that slides over the entire input and extracts features that can then be trained using either quantum circuits or classical layers. On the other hand, the QCNN downsizes the input to match the first circuit's qubit count and performs pooling to reduce the number of qubits in successive circuits until reaching the desired output size. 

We systematically evaluate these HQNN models under different noise conditions\footnote{We consider different the noise conditions that take into account all the types of noise that can occur in NISQ devices, i.e., decoherence (modeled by amplitude damping, depolarization, and phase damping), gate errors (modeled by bit flip and phase flip), and readout errors (modeled by depolarization).} to assess their robustness and practical applicability in NISQ devices. In our experiments, we embed the noise in the HQNNs during training to emulate the HQNNs' learning process on noisy NISQ devices. Our findings contribute to the development of noise-resilient QML algorithms and provide insights into whether HQNNs can outperform their classical model counterparts in practical real-world scenarios.



\subsection{Motivational Analysis}

\begin{figure}[ht]
  \centering
  \includegraphics[width=\linewidth]{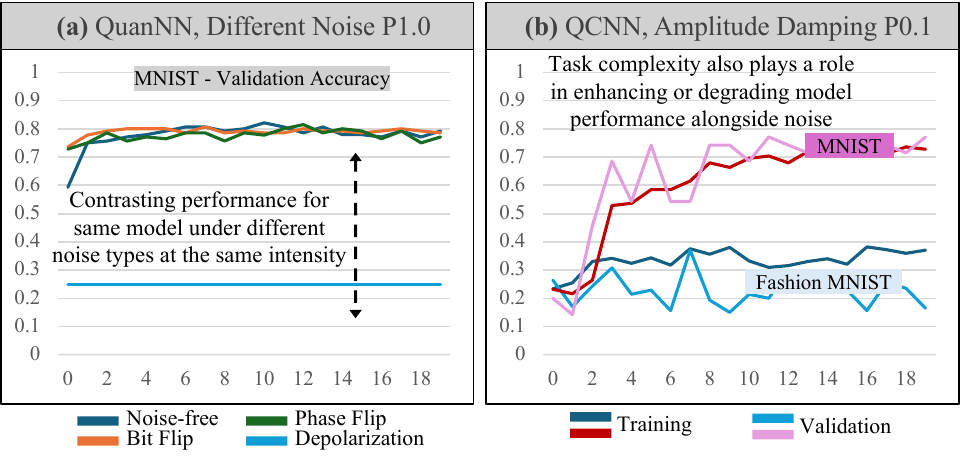}
  \caption{Performance variations of HQNNs under different noise types and task complexities. (a) QuanNN's noise robustness under different noise channels with probability 1.0, for the MNIST dataset. (b) QCNN's noise robustness under Amplitude Damping noise with probability 0.1, for the MNIST and Fashion MNIST datasets.}
  \label{fig:motivation}
\end{figure}

A detailed evaluation of the performance of HQNN under different types and intensities of noise on different ML tasks is crucial for practical development. Fig.~\ref{fig:motivation} highlights key challenges and insights that motivate this study.

Our experimental results reveal that, despite identical circuit architectures, model performance varies significantly under different conditions. In Fig.~\ref{fig:motivation}(a), we observe that the QuanNN trained on the MNIST dataset exhibits diverse behaviors at high noise probabilities (100\%) across different noise channels. Notably, the Phase Flip noise with a probability of 1.0 improves the model's robustness, even outperforming the noise-free model baseline. In contrast, the depolarization noise at the same probability results in a model that entirely fails to learn. This suggests a non-trivial relationship between noise types and model performance.

Moreover, task complexity plays a crucial role in model performance. Fig.~\ref{fig:motivation}(b) illustrates this by comparing the QCNN performance on a subset of four classes of the MNIST and Fashion MNIST datasets, with Amplitude Damping noise at 0.1 probability. The relatively simpler MNIST classification task yields better performance compared to the slightly more complex Fashion MNIST, which shows noticeable performance degradation. These observations underscore the importance of considering both noise characteristics and tasks when developing robust HQNN models. Therefore, a comprehensive analysis is essential to understand the intricate interplay between these factors and optimize HQNN performance across various real-world quantum computing scenarios.

\subsection{Our Contributions}
Our key contributions are summarized below: 
\begin{itemize}[leftmargin=*]

\item  \textbf{Systematic Noise Robustness Analysis}
We evaluate HQNN performance, namely QCNN and QuanNN, under different noise conditions by independently injecting five different quantum noise channels (Bit Flip, Phase Flip, Phase Damping, Amplitude Damping, and Depolarizing Noise) into the quantum circuits, with noise probabilities ranging from $0.1$ to $1.0$. The noise is introduced after each parametric gate and entanglement block, simulating gate errors, decoherence, and readout errors. \textit{This systematic approach reveals how different noise sources impact performance across varying probabilities.}

\item \textbf{Scalability to Larger Circuits and Diverse Tasks}
Our study extends previous 2-qubit noise analyses~\cite{kashif2024investigating} by employing 4-qubit circuit architectures, analyzing the noise robustness for two distinct image classification tasks, MNIST and Fashion MNIST\footnote{Note that the MNIST and Fashion MNIST datasets serve as widely recognized benchmarks for QML research. The constraints of current NISQ devices limit the implementation of HQNNs to small-scale quantum circuits.}. \textit{This setup enables a deeper understanding of how task demands and circuit complexity influence noise resilience}.

\item \textbf{Identifying Noise-Resilient Architectures} Through comparative evaluation of QuanNN and QCNN, we aim to identify which architecture demonstrates greater resilience to specific noise types across varying probability levels and datasets. \textit{This analysis will guide the development of robust HQNN designs for diverse noise conditions and task complexities}.

\item  \textbf{Guidelines for HQNN Design in NISQ Environments} Our comprehensive analysis of how noise impacts HQNN architectures provides valuable insights for designing noise-aware HQNN models. By demonstrating how various models respond to specific noise channels and intensities, we lay the groundwork for developing more resilient HQNNs.
\end{itemize}

Our findings highlight the importance of considering noise characteristics when selecting or designing quantum circuits for specific tasks, contributing to the development of robust HQNNs suitable for NISQ-era devices and guiding future research in noise mitigation strategies for HQNNs.

\textbf{Paper Organization:} \Cref{sec:background} provides details on the HQNN algorithms and noise gates used in this study. \Cref{sec:framework} discusses all the steps of our proposed noise robustness analysis framework. \Cref{sec:results} reports the experimental results of our framework for various benchmarks, noise models, and intensities. \Cref{sec:conclusion} concludes the paper.

\section{Background}
\label{sec:background}

\subsection{HQNN Algorithms}

HQNNs leverage quantum circuits to enhance classical deep learning models, particularly in tasks like image classification. We focus on Quanvolutional Neural Networks (QuanNNs) and Quantum Convolutional Neural Networks (QCNNs) due to their widespread adoption and direct applicability to vision-based tasks.

\subsubsection{Quanvolutional Neural Network (QuanNN)}
The Quanvolutional Neural Network (QuanNN)~\cite{QuanNN:2019} is a state-of-the-art hybrid quantum-classical architecture that integrates quantum filters for feature extraction, called \textit{quanvolutional filters}, into a classical CNN pipeline, as shown in Fig.~\ref{fig:hqnn}. Each quanvolutional filter operates by convolving over the input features, and its size depends on the number of qubits used\footnote{In our case, using 4 qubits results in a 2$\times$2 grid filter.}. The QuanNN operates through the following steps:

\begin{enumerate}[leftmargin=*]
    \item Embedding local subsections of input data into quantum states through quanvolutional filters.
    \item Processing these states through a quantum circuit with trainable parameters (unitary U).
    \item Measuring the outcome and post-processing it through classical layers.
    \item Iterating the same procedure over different regions of the input to construct a multi-channel output feature map.
\end{enumerate}

\subsubsection{Quantum Convolutional Neural Networks (QCNN)}

\begin{figure}[ht]
  \centering
  \includegraphics[width=.8\linewidth]{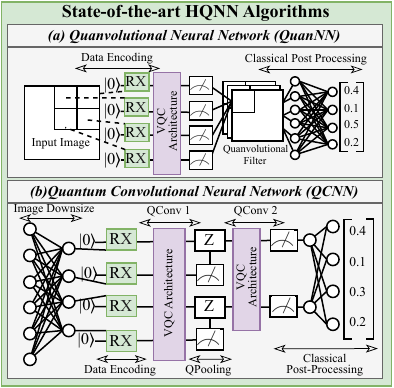}
  \caption{Circuit structure and data process pipeline of our chosen HQNN algorithm. (a) The Quanvolutional Neural Network has a quantum filter that convolves over the entire image. (b) The Quantum Convolutional Neural Network uses classical layers to downsize the image.}
  \label{fig:hqnn}
\end{figure}

The Quantum Convolutional Neural Network (QCNN)~\cite{QCNN:2019} is an advanced HQNN algorithm inspired by classical CNNs but implemented using VQCs. Unlike the QuanNN, it features fully quantum-based convolutional and pooling layers. The QCNN is notable for its logarithmic scaling of variational parameters, requiring only $O(\log(N))$ trainable parameters for input sizes of $N$ qubits, which makes it suitable for implementation on NISQ devices. As shown in Fig.~\ref{fig:hqnn}, its architecture consists of an input encoding circuit layer, quantum convolutional and pooling layers, and a circuit measurement layer, with all but the last layer containing parametric quantum gates. The QCNN is considered an HQNN as it uses classical optimization techniques to optimize the parametric gate weights. Although the quantum layers in QCNNs efficiently extract features through qubit interactions, its scalability remains constrained by the limited qubit availability in NISQ devices. To address this, our implementation incorporates single classical convolution and pooling layers to pre-process high-dimensional inputs, ensuring compatibility with quantum circuits while preserving essential features.

\subsection{Quantum Noise \& Error Gates}
NISQ-era quantum devices are inherently prone to noise, introducing errors that degrade HQNN performance at different computational stages. These undesired random fluctuations arise from environmental interference and fundamental quantum mechanical principles, making quantum noise inherently probabilistic and more detrimental than classical noise. The following sections detail the quantum noise types used in this work.


\subsubsection{Bit Flip}
A bit flip error models classical bit errors in quantum computing, causing a qubit to switch between states $\ket{0}$ and $\ket{1}$ with probability $p$. This error, also known as a Pauli X error, can be represented as in Equation~\ref{eq:bitflip}, using Kraus matrices with a probability $p \in [0, 1]$. 
\begin{equation}
K_0 = \sqrt{1-p}I, \quad K_1 = \sqrt{p}X
\label{eq:bitflip}
\end{equation}
where $I$ is the identity matrix and $X$ is the Pauli X matrix.

\vspace{5pt}
\subsubsection{Phase Flip}
Unlike classical bits, qubits possess a phase component representing a rotation around the Z-axis. A phase flip error in quantum computing affects the relative phase of a quantum state without altering its amplitude. This operation, represented by the Pauli Z gate, leaves the $\ket{0}$ state unchanged while multiplying the $\ket{1}$ state by $-1$. The phase flip error can be described using Kraus matrices with a probability $p \in [0, 1]$, as in Equation~\ref{eq:phaseflip}. 
\begin{equation}
K_0 = \sqrt{1-p}I, \quad K_1 = \sqrt{p}Z
\label{eq:phaseflip}
\end{equation}
where $I$ is the identity matrix, and $Z$ is the Pauli Z matrix.

\vspace{5pt}
\subsubsection{Depolarization Channel}
The depolarizing channel is a comprehensive noise model in quantum computing that combines aspects of bit flip and phase flip errors. Unlike specific errors affecting particular qubit aspects, depolarizing errors are non-selective, potentially randomizing a qubit's state to any point on the Bloch sphere with a given probability.
In this process, the qubit state may be replaced by a completely mixed state $\frac{I}{2}$ with probability $p$, reducing the quantum state's purity and ``smearing'' its Bloch sphere representation towards the center. As $p$ increases, more quantum information and coherence are lost. The depolarizing channel can be represented using Kraus matrices as in Equation~\ref{eq:depolarize}. 
\begin{equation}
\begin{adjustbox}{max width=.9\linewidth}
$K_0 = \sqrt{1-p}I, \quad K_1 = \sqrt{\frac{p}{3}}X, \quad  K_2 = \sqrt{\frac{p}{3}}Y, \quad  K_3 = \sqrt{\frac{p}{3}}Z$
\end{adjustbox}
\label{eq:depolarize}
\end{equation}
where $p \in [0, 1]$ is the depolarization probability, equally distributed among X, Y, and Z Pauli operations.

\vspace{5pt}
\subsubsection{Phase Damping}
Phase damping, or dephasing, is a quantum noise process that causes a qubit to lose its quantum coherence over time without changing its energy. It gradually erases the phase relationship between $\ket{0}$ and $\ket{1}$ states, transforming pure quantum states into mixed states, affecting phase information. This process does not alter the probabilities of measuring $\ket{0}$ or $\ket{1}$, but effectively turns quantum behavior into classical behavior, impacting the qubit's ability to maintain superposition and perform quantum computations. The error can be represented as in Equation~\ref{eq:phasedamp}, using Kraus matrices with probability $\gamma \in [0, 1]$. 
\begin{equation}
K_0 = \begin{pmatrix} 1 & 0 \\ 0 & \sqrt{1-\gamma} \end{pmatrix}, \quad 
K_1 = \begin{pmatrix} 0 & 0 \\ 0 & \sqrt{\gamma} \end{pmatrix}
\label{eq:phasedamp}
\end{equation}

\vspace{5pt}
\subsubsection{Amplitude Damping}
Amplitude damping represents energy loss in a quantum system, causing transitions from excited state $\ket{1}$ to ground state $\ket{0}$. It occurs due to interactions with the environment, leading to gradual energy dissipation. Unlike phase damping, amplitude damping involves state transitions. While phase damping only affects the relative phase between qubit states, amplitude damping involves the probability of a qubit transitioning from the excited state $\ket{1}$ to the ground state $\ket{0}$. The Kraus operators for amplitude damping with probability $\gamma \in [0, 1]$ can be represented as in Equation~\ref{eq:ampdamp}.
\begin{equation}
K_0 = \begin{pmatrix} 1 & 0 \\ 0 & \sqrt{1-\gamma} \end{pmatrix}, \quad 
K_1 = \begin{pmatrix} 0 & \sqrt{\gamma} \\ 0 & 0 \end{pmatrix}
\label{eq:ampdamp}
\end{equation}

\subsection{Related Work}

Recent advancements in HQNNs have demonstrated their potential in combining classical deep learning and quantum computing to conduct classification tasks. Zaman et al.~\cite{zaman2024comparative} conducted a comparative analysis of three HQNN models (QuanNN, QCNN, and Quantum Residual Networks), evaluating their performance under different circuit architectures using the MNIST dataset. While their work provides valuable insights into HQNN behavior, it assumes an ideal noise-free environment and does not consider real-world noise in NISQ devices.

Building upon this foundation, Kashif et al.~\cite{kashif2024investigating} investigated the influence of five noise channels (amplitude damping, bit flip, depolarization, phase flip, and phase damping) on a two-qubit Variational Quantum Circuit (VQC) architecture for binary classification on MNIST. Similarly, Domingo et al.~\cite{reservoir_qc} explored noise robustness in quantum reservoir computing, demonstrating that specific noise types can be either detrimental or beneficial, depending on the learning task.

Ahmed et al.~\cite{ahmed2025quantumneuralnetworkscomparative} examined the impact of readout errors in HQNNs by injecting noise before measurement. In a related study, Escudero et al.~\cite{qnn_motten} examined the Mottonen state preparation algorithm under various noise models and evaluated the degradation of quantum states due to noise propagation through multiple HQNN layers, however they don't specify the exact noise channels used in their implementation.

As shown in \Cref{tab:related_works}, our work extends these prior efforts in several key ways. We adopt the best-performing architectures of QuanNN and QCNN from Zaman et al.~\cite{zaman2024comparative} and examine their robustness under different noise channels. Unlike previous studies that used simplified binary classification tasks, we employ 4-qubit quantum circuits for various multi-classification tasks using MNIST and Fashion-MNIST.

\begin{table}[!ht]
    \centering
    \caption{Comparison with related works.}
    \label{tab:related_works}
    \begin{adjustbox}{max width=\linewidth}
    \begin{tabular}{|c|c|c|c|c|c|}
        \hline
            \textbf{Reference} & \textbf{\makecell{Multi\\ Class}} & \textbf{\makecell{Decoherence\\ }} & \textbf{\makecell{Readout\\ Error}} & \textbf{\makecell{Gate\\ Error}} & \textbf{\makecell{Multi\\ Datasets}}\\ 
        \hline
        Zaman et al.~\cite{zaman2024comparative} & \checkmark &&&&\\ 
        \hline
        Ahmed et al.~\cite{ahmed2025quantumneuralnetworkscomparative} & \checkmark && \checkmark &&\\ 
        \hline
        Kashif et al.~\cite{kashif2024investigating} && \checkmark & \checkmark & \checkmark &\\ 
        \hline
        Domingo et al.~\cite{reservoir_qc} && \checkmark && \checkmark &\\ 
        \hline
        Escudero et al.~\cite{qnn_motten} &\checkmark &\checkmark &&\checkmark &\\ 
        \hline
        \textbf{This work} & \checkmark & \checkmark & \checkmark & \checkmark & \checkmark\\ 
        \hline
    \end{tabular}
    \end{adjustbox}
\end{table}

Unlike previous studies that only consider readout errors~\cite{ahmed2025quantumneuralnetworkscomparative}, we systematically inject noise in every layer of the quantum circuit along with measurement noise, to simulate gate errors, decoherence, and readout errors, and to analyze their combined impact. Lastly, in contrast to Escudero et al.'s approach~\cite{qnn_motten} that focuses on state preparation, our method directly encodes image data into the parameters of RY gates in the encoding layer, followed by specific quantum layer design in both QuanNN and QCNN architectures. This ensures a consistent encoding framework while enabling a systematic analysis of noise effects across different circuit architectures for image classification tasks.

\section{Noise Robustness Analysis Framework}
\label{sec:framework}

\begin{figure*}[h]
    \centering
    \includegraphics[width=\linewidth]{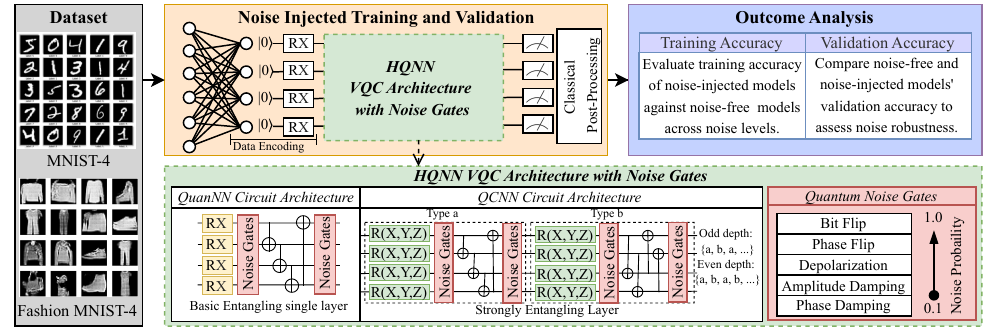}
    \caption{\footnotesize Our methodology. A comprehensive comparative analysis of two different HQNN models (QuanNN and QCNN) is performed with different noise gate types (bit flip, phase flip, depolarization, amplitude damping, and phase damping) with different intensities, for MNIST and Fashion MNIST datasets. The odd and even depth in the strongly entangling configuration of the QCNN denotes how the layer is repeated when the number of layers are increased. The evaluation metrics used for all the experiments are training and validation accuracy.}
    \label{fig:methodology}
\end{figure*}

This study analyzes two HQNN algorithms under five distinct quantum noise models, evaluating their performance across varying noise levels. By experimenting with different noise intensities, we examine the impact on HQNN models' behavior and performance across diverse machine learning tasks. \Cref{fig:methodology} provides an outline of our methodology.  

\subsection{Specifications}
\subsubsection{Dataset}
For training and validation of the HQNN models, we utilize subsets of the MNIST\cite{MNIST} and Fashion-MNIST\cite{FMNIST} datasets\footnote{For both datasets, our analysis is restricted to the first four classes to ensure compatibility between the number of classes and the four-qubit quantum circuits in our experiments.}. The MNIST dataset is widely adopted in QML due to its simplicity and established benchmark status in image classification, making it ideal for evaluating noise effects on HQNNs. In contrast, Fashion-MNIST, while maintaining the same image size and structure as MNIST, presents a more complex classification challenge. This increased complexity is due to its greater intraclass variability and features derived from downsampled color photographs\cite{FMNIST}. These characteristics introduce subtle distinctions between clothing categories that are more challenging to distinguish compared to the clear-cut digit shapes in MNIST \cite{FMNIST}. Conducting experiments on both datasets allows us to investigate whether noise robustness in HQNNs is influenced by different task demands.

\subsubsection{Selected HQNN Algorithms}
For this study, we selected the QuanNN and QCNN models due to their mainstream usage in QML for image classification. The suitability for this study is supported by prior work in~\cite{zaman2024comparative}, which evaluated these models in various circuit architectures, qubit counts, and layer configurations, demonstrating their effectiveness and adaptability in QML classification tasks.

Both HQNN architectures in our study incorporate a classical layer immediately following the quantum layers to convert quantum measurements into classical probabilities. The QCNN architecture additionally employs a classical pre-processing layer to reduce the input dimension prior to quantum encoding, distinguishing it from the QuanNN design.

\subsection{Noisy Training \& Validation}
We investigate the performance of the QuanNN and QCNN for image classification under various noise conditions, by adopting the best-performing configurations identified in~\cite{zaman2024comparative}. Using the experimental setup and hyperparameters as described in Table~\ref{tab:exp_setup}, we first trained both models without noise to validate their noise-free performance. Afterward, we implemented noise injection during training and validation to simulate noisy conditions.

\subsubsection{Circuit Architecture \& Noise-Injection}
The HQNN configurations consist of the QuanNN with 3 basic entangling layers, and the QCNN with 3 strongly entangling layers, as illustrated in \Cref{fig:methodology}.

To establish a baseline and validate our model selection, we first reproduced the experiments without noise. Both QuanNN and QCNN architectures achieved high performance, with a validation accuracy of approximately 80\%, confirming their effectiveness for the given tasks.

To comprehensively simulate the noisy behavior of NISQ devices, we introduced noise gates after every parametric gate block and entanglement block. In our experiments, we inject noise using the PennyLane library, which emulates decoherence and noise errors that occur in real quantum systems. \Cref{fig:methodology} shows the placement of these noise channels, enabling a systematic evaluation of the impact of various noise types on HQNNs' performance.

\begin{figure*}[h]
    \centering
    \includegraphics[width=\linewidth]{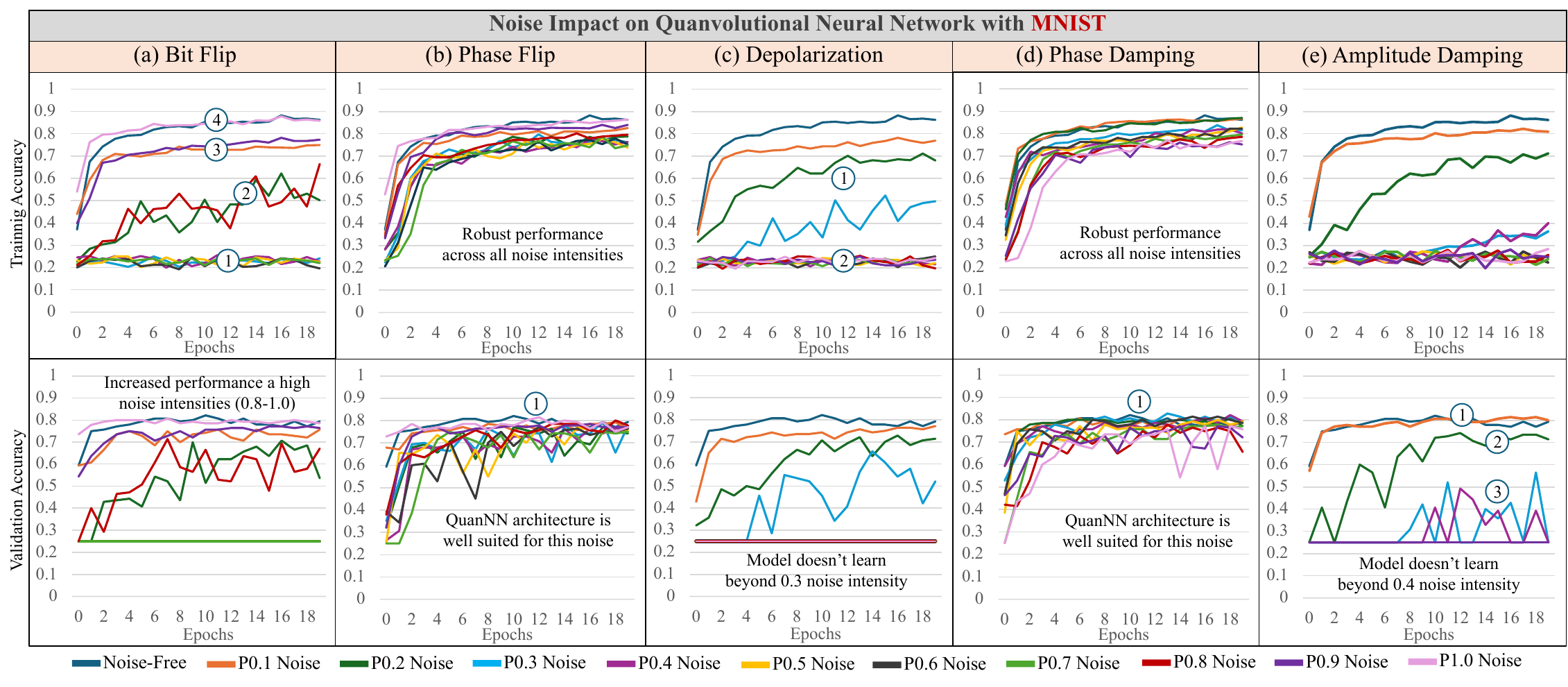}
    \vspace{-20pt}
    \caption{Comparative performance analysis of the QuanNN on the MNIST dataset under various noise channels with 0.1 to 1.0 intensities, benchmarked against the noise-free model performance. The QuanNN shows robust performance for phase damping and phase flip across all noise levels. For other noise channels, its performance degrades with increasing noise intensity, except for the bit flip noise, for which the QuanNN demonstrates high robustness at the 1.0 level.}
    \label{fig:mnist_quan}
\end{figure*}

\subsubsection{Quantum Noise Gates}
We examine the impact of five types of quantum noise commonly encountered in NISQ devices: Bit Flip, Phase Flip, Depolarization, Phase Damping, and Amplitude Damping. We implemented these noise gates using the PennyLane library because of its accessibility and widespread use in QML research. To simulate varying degrees of noise, we assign probability values to each noise gate, representing the likelihood of its occurrence in the circuit. These probabilities, ranging from 0.1 to 1.0 with increments of 0.1, result in ten probability values per noise gate type. 

\subsection{Outcome}
To assess the robustness of the HQNN models against noise, our analysis focuses on two key performance metrics: training and validation accuracies. We conducted a thorough comparison of these metrics across all noise channels with varying intensities, enabling us to:
\begin{itemize}[leftmargin=*]
    \item Assess the HQNN models' resilience across a spectrum of noise intensities.
    \item Identify potential thresholds where performance significantly degrades.
    \item Examine potential non-linear correlations between noise probability and model accuracy.
\end{itemize}

This analysis provides a comprehensive understanding of how different types and levels of quantum noise impact the performance of QuanNN and QCNN models. This evaluation offers insights into the practical applicability of HQNN algorithms in NISQ environments and guides future research on noise-resilient QML.


\section{Results and Discussions}
\label{sec:results}

In this section, we present a comprehensive robustness analysis of QuanNN and QCNN architectures against various types of quantum noise and errors on two distinct machine learning tasks: MNIST and Fashion MNIST classification. 


\subsection{Classical Optimization Environment and Setup}

In our experimental setup, we maintained a uniform classical optimization environment for both classical and quantum components, as outlined in Table~\ref{tab:exp_setup}. The HQNN architectures in our study have a classical layer immediately following the quantum layers, to convert the quantum measurements into classical probabilities. However, the QCNN architecture additionally employs a classical pre-processing layer to downsize the input dimension prior to quantum encoding. 

All quantum layers are implemented in PennyLane for seamless integration with PyTorch, facilitating end-to-end gradient-based optimization of quantum gate parameters\footnote{https://docs.pennylane.ai/en/stable/introduction/interfaces/torch.html}. The noise gates are also implemented in PennyLane to facilitate their integration with the rest of the quantum circuit\footnote{https://pennylane.ai/blog/2021/05/how-to-simulate-noise-with-pennylane}. All the hyperparameters used are listed in Table~\ref{tab:exp_setup}. Note that we employ dataset-specific learning rates that are optimized separately for MNIST and Fashion-MNIST datasets. We determined these learning rate values of 0.01 for the MNIST and 0.005 for the Fashion MNIST dataset through preliminary experiments with noise-free models, ensuring a consistent baseline for fair comparison with noise-injected models across different conditions.

\begin{table}[!t]
    \centering
    \caption{Training Environment Setup Details}
    \label{tab:exp_setup}
    \begin{tabular}{|c|c|}
        \hline
            \textbf{Components} & \textbf{Details} \\ 
        \hline
            Software Frame Work & \verb|PennyLane|\cite{pennylane_diff} \\
        \hline
            Back-End Simulator & \verb|lightning.qubit| \\ 
        \hline
            Noise Simulator & \verb|default.mixed| \\
        \hline
            Back-End Machine & \verb| NVIDIA RTX 6000 Ada | \\ 
        \hline
            Deep-Learning Interface & \verb|Pytorch| \\ 
        \hline
            Data-set & \verb|MNIST|~\cite{MNIST}, \verb|Fashion-MNIST|\cite{FMNIST} \\  
        \hline
            Training \& Testing Samples &  \verb|500|, \verb|150| \\
        \hline
            Epoch, Batch-Size, LR & \verb|20|, \verb|5|, \verb|0.01 & 0.005| \\ 
        \hline
    \end{tabular}
\end{table}



\subsection{Noise Robustness Analysis of QuanNN}
This section presents the resilience of QuanNNs against various quantum noise types. 

\paragraph{QuanNN robustness against Bit Flip Noise}

\begin{figure*}[h]
    \centering
    \includegraphics[width=\linewidth]{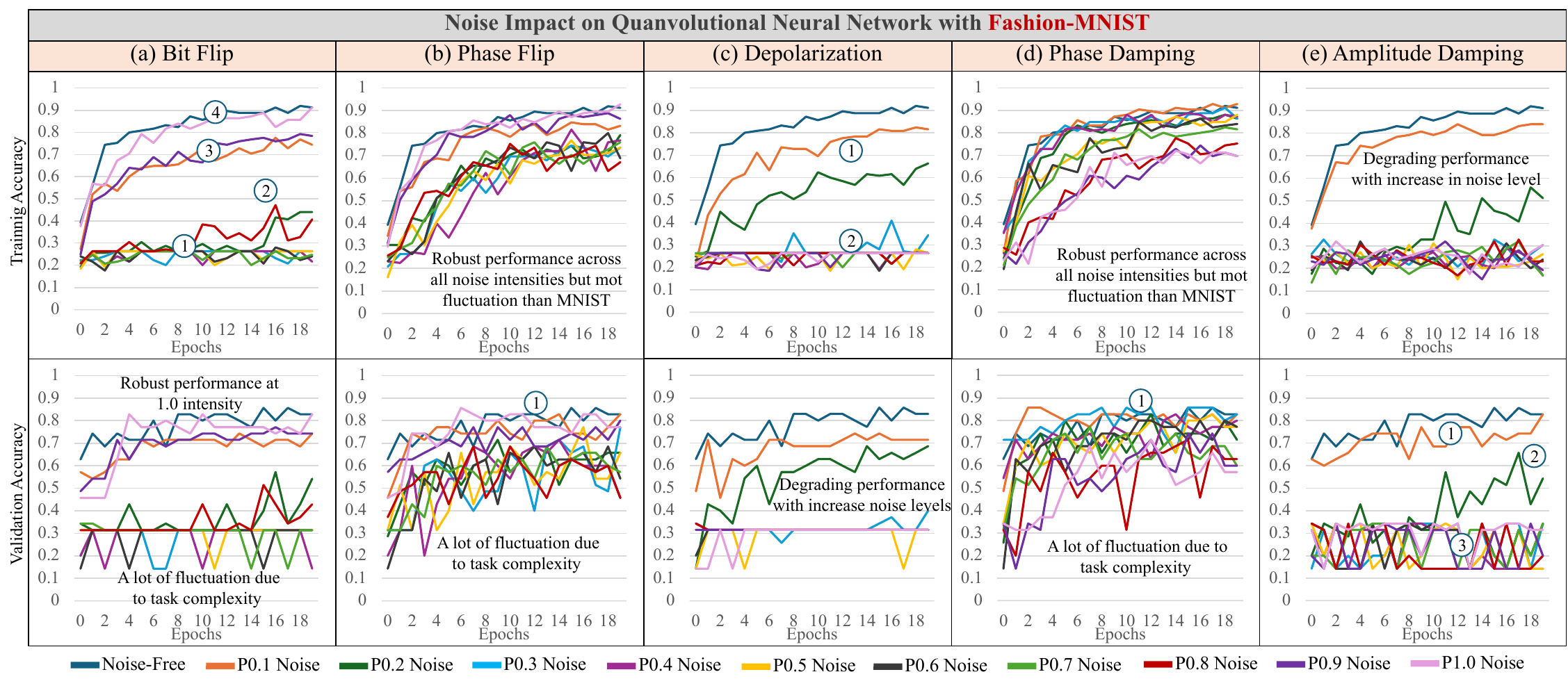}
    \vspace{-20pt}
    \caption{Comparative performance analysis of the QuanNN on the Fashion-MNIST dataset under various noise channels with 0.1 to 1.0 intensities, benchmarked against the noise-free model performance. The model shows great robustness to phase-related noise but performance degradation with other noise channels as intensity increases. For bit flip noise at 1.0 level, the model shows great robustness. The performance is more fluctuating compared to MNIST.}
    \label{fig:fmnist_quan}
\end{figure*}

Figures~\ref{fig:mnist_quan}(a) and~\ref{fig:fmnist_quan}(a) illustrate the comparative analysis of QuanNN performance under noise-free conditions and Bit Flip noise for MNIST and Fashion-MNIST respectively, revealing intriguing patterns in the model response across both datasets. At noise probabilities between 0.3 and 0.7 (pointer~\rpoint{1} in Fig.~\ref{fig:mnist_quan}(a) \& ~\ref{fig:fmnist_quan}(a)), the QuanNN training accuracy drops below 30\% for both datasets, indicating significant performance degradation. However, for noise probabilities of 0.2 and 0.8 (pointer~\rpoint{2} in Fig.~\ref{fig:mnist_quan}(a)), the model shows signs of adaptation, with the training accuracy gradually rising from 30\% to 60\% for the MNIST dataset, indicating an increasing robustness to Bit-Flip noise. However, this ascent is slower for Fashion MNIST at noise probabilities of 0.2 and 0.8 (pointer~\rpoint{2} in Fig.~\ref{fig:fmnist_quan}(a)), reaching a maximum validation accuracy of only 50\%. At noise probabilities 0.1 and 0.9, the accuracy stabilizes around 75\% for both datasets (pointer~\rpoint{3}).

Notably, for noise probability of 1.0, the model exhibits near noise-free training accuracy for both MNIST and Fashion MNIST (pointer~\rpoint{4}), suggesting that the consistent application of X gates with 100\% Bit Flip probability allows the model to adapt and learn to compensate for these deterministic transformations, resulting in minimal performance impact. The validation accuracy follows the same trend, confirming the model's ability to adapt to this noise and generalize for unseen data. These results highlight critical thresholds in Bit Flip noise tolerance, beyond which QuanNN's performance degrades significantly. 

\paragraph{QuanNN robustness against Phase Flip Noise}

Figures~\ref{fig:mnist_quan}(b) and~\ref{fig:fmnist_quan}(b) illustrate the impact on the QuanNN's robustness against phase flip noise for MNIST and Fashion MNIST, respectively. The observed results demonstrate that the QuanNN remains considerably robust to the phase flip noise across all probabilities. This resilience is noteworthy since the phase flip noise is a common quantum error, equivalent to applying a Z gate, which causes a sign change in the quantum state's phase. The QuanNN's robust performance is likely attributed to the inherent insensitivity of Z-basis measurements to phase flips, as these errors do not alter measurement outcome probabilities. Moreover, phase flip noise may serve as an implicit regularization, enhancing the model's generalization capabilities. 

While the validation accuracy for MNIST remains stable across all noise probabilities (pointer~\rpoint{1} in Fig.~\ref{fig:mnist_quan}(b)), the validation accuracy for Fashion MNIST exhibits more fluctuations (pointer~\rpoint{1} in Fig.~\ref{fig:fmnist_quan}(b)), likely due to the performance variability introduced in a more complex task. However, despite these fluctuations, the overall trend confirms that the QuanNN's performance is minimally affected by phase flip noise, in contrast to its sensitivity to other forms of quantum noise such as depolarization and amplitude damping.

\paragraph{QuanNN robustness against Depolarization Noise}

The performance of QuanNN under depolarization noise is depicted in Fig.~\ref{fig:mnist_quan}(c) for MNIST and Fig.~\ref{fig:fmnist_quan}(c) for Fashion MNIST, illustrating a clear correlation between noise level and model performance. At low noise levels (0.1-0.2 for Fashion MNIST and up to 0.3 for MNIST, pointer~\rpoint{1} in Fig.~\ref{fig:mnist_quan}(c) and Fig.~\ref{fig:fmnist_quan}(c)), the QuanNN maintains a gradual accuracy improvement, showing an ability to learn and adapt to minimal noise. However, for noise levels $\geq$ 0.3 for Fashion MNIST (pointer~\rpoint{2}) or $\geq$ 0.4 for MNIST (pointer~\rpoint{2}), the training accuracy degrades significantly, indicating that the model fails to learn effectively. This behavior can be attributed to the nature of the depolarization channel, which combines both phase flip and bit flip errors. As noise levels increase, this combination of errors introduces excessive randomness into the quantum state, which deteriorates the information encoded in the qubits. Hence, depolarization disrupts meaningful pattern extraction and critically affects the QuanNN's learning capabilities beyond moderate noise levels.

\paragraph{QuanNN robustness against Phase Damping Noise}

The performance of QuanNN when subjected to phase damping noise is presented in Fig.~\ref{fig:mnist_quan}(d) for MNIST and Fig.~\ref{fig:fmnist_quan}(d) for Fashion MNIST. Similar to its behavior under phase flip noise, the QuanNN demonstrates robust performance across all intensities of phase damping noise. The validation accuracy across all noise levels closely aligns with the noise-free performance for MNIST (pointer~\rpoint{1} in Fig.~\ref{fig:mnist_quan}(d)), demonstrating that the phase damping does not significantly impact the QuanNN's learning process. 
While the validation accuracy of the QuanNN under phase damping noise exhibits some variability for the Fashion MNIST dataset (pointer~\rpoint{1} in Fig.~\ref{fig:fmnist_quan}(d)), the QuanNN maintains consistent performance across all noise levels for the MNIST dataset, showcasing remarkable robustness. 

Given that, in typical quantum systems, the phase damping noise causes information loss without energy loss and potentially degrades the performance, the QuanNN's insensitivity suggests that its architecture is inherently adapting to phase-related disturbances during training, or that phase damping has minimal impact on its decision-making process. 
\begin{figure*}[h]
    \centering
    \includegraphics[width=\linewidth]{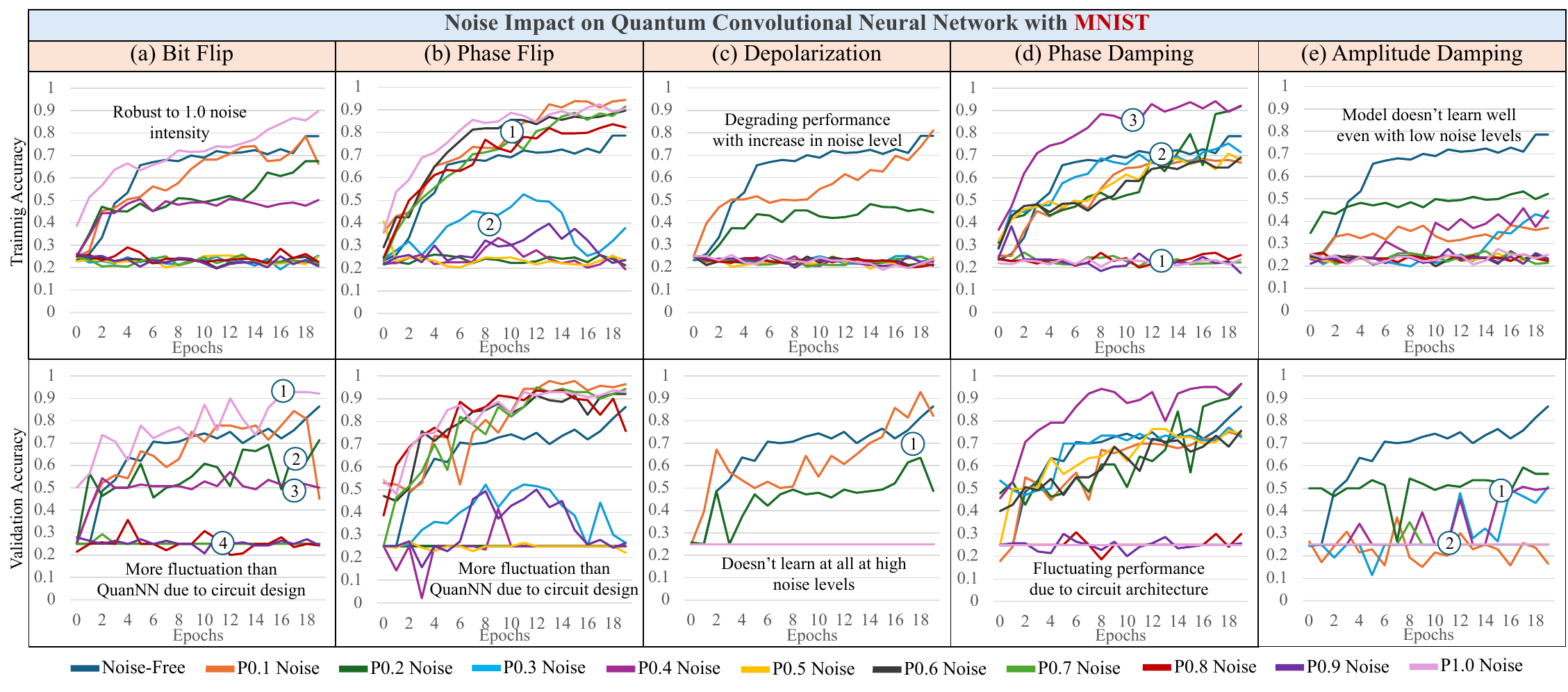}
    \vspace{-20pt}
    \caption{Comparative performance analysis of the QCNN on the MNIST dataset under various noise channels with 0.1 to 1.0 intensities, benchmarked against the noise-free model performance. The QCNN model shows significant performance degradation with increasing intensity for depolarization, but exhibits fluctuating performance with no consistent trends as noise intensity increases for other noise types.}
    \label{fig:mnist_qcnn}
\end{figure*}

\paragraph{QuanNN robustness against Amplitude Damping}

The training and validation results of the QuanNN under amplitude damping noise, as depicted in Fig.~\ref{fig:mnist_quan}(e) for MNIST and the Fig.~\ref{fig:fmnist_quan}(e) for Fashion MNIST, mirror the behavior under depolarization noise. The QuanNN's performance degrades as the noise level increases, demonstrating the significant impact of amplitude damping on the quantum state.

At a noise level of 0.1 (pointer~\rpoint{1} in Fig.~\ref{fig:mnist_quan}(e)), the validation accuracy remains close to the noise-free baseline, indicating minimal impact on performance. However, for a noise level of 0.2 (pointer~\rpoint{2}), the accuracy drops to approximately 75\%, showing a slight degradation compared to the noise-free scenario. 

Beyond 0.3 , the QuanNN's accuracy declines sharply, particularly for the Fashion MNIST dataset (pointer~\rpoint{3} in Fig.~\ref{fig:fmnist_quan}(e)). For the MNIST (pointer~\rpoint{3} in Fig.~\ref{fig:mnist_quan}(e)), the amplitude damping with noise levels of 0.3 and 0.4 causes fluctuations between 25-50\% , suggesting some residual learning capability, but the performance collapses for noise levels of 0.5 or higher. This behavior can be explained by looking at the nature of amplitude damping in quantum physics. Since the amplitude damping noise models energy dissipation in quantum systems, high noise probabilities reduce the state space available for computation, thus limiting the model's ability to learn and generalize from the data. 


\subsection{Noise Robustness Analysis of QCNN}
This section evaluate the resilience of QCNNs against various types of quantum noise. 

\begin{figure*}[h]
    \centering
    \includegraphics[width=\linewidth]{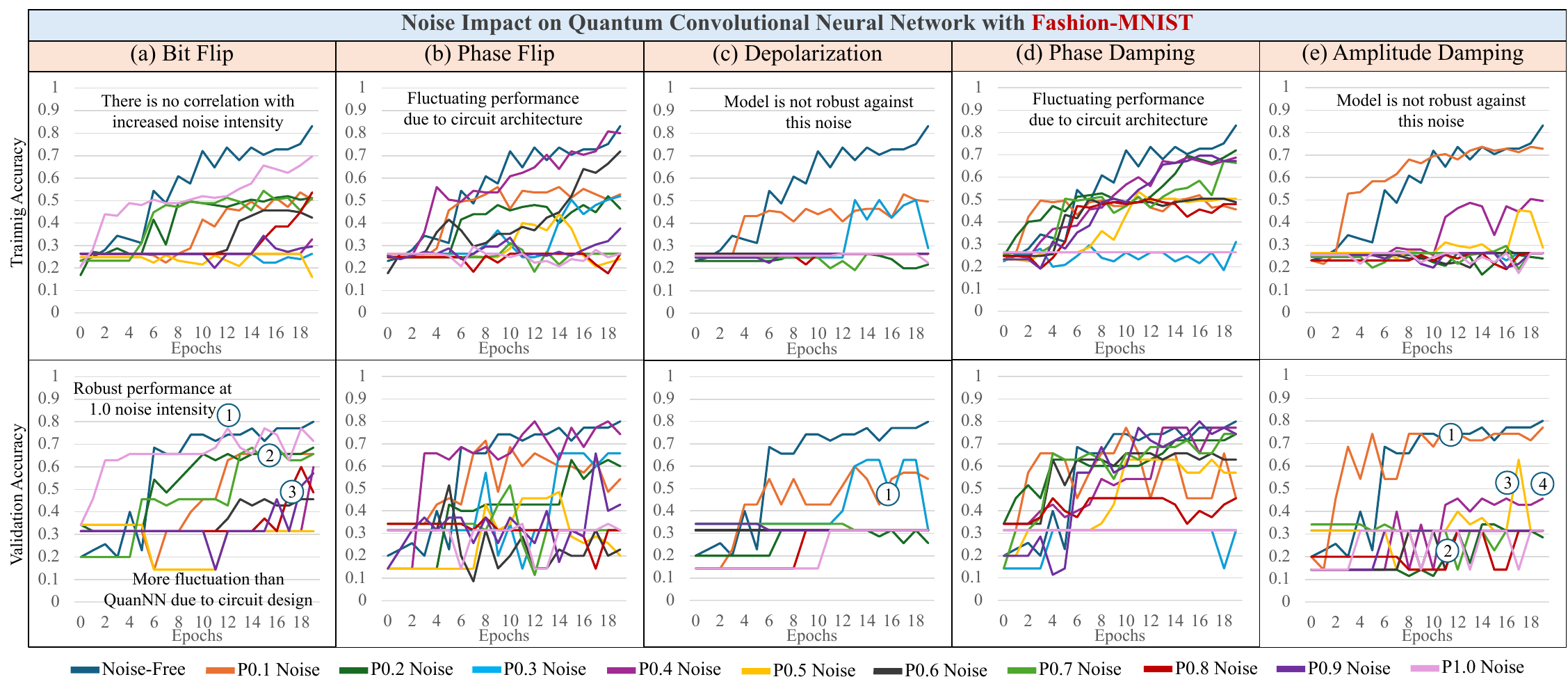}
    \vspace{-20pt}
    \caption{Comparative performance analysis of the QCNN on the Fashion-MNIST dataset under various noise channels with 0.1 to 1.0 intensities, benchmarked against the noise-free model performance. The QCNN shows high fluctuations and no clear trends in performance in relation to noise level.}
    \label{fig:fmnist_qcnn}
\end{figure*}

\paragraph{QCNN robustness against Bit Flip Noise}
The performance of QCNNs under bit flip noise is presented in Fig.~\ref{fig:mnist_qcnn}(a) for MNIST and Fig.~\ref{fig:fmnist_qcnn}(a) for Fashion MNIST, revealing intriguing patterns across noise levels and datasets. For MNIST, at 1.0 noise probability (pointer~\rpoint{1} in Fig.~\ref{fig:mnist_qcnn}(a)), the QCNN outperforms the noise-free model, demonstrating effective noise adaptation to the deterministic behavior of the noise gates, similar to the QuanNN's response observed in \Cref{fig:mnist_quan}. For noise levels of 0.2 (pointer~\rpoint{2}) and 0.4 (pointer~\rpoint{3}), the QCNN shows gradual improvement, reaching accuracies of 70\% and 50\%, respectively. However, for noise levels 0.3 and between 0.4 and 0.9 (pointer~\rpoint{4}), the model's accuracy flattens at around 25\%..

The QCNN for the Fashion MNIST exhibits greater performance fluctuations than for the MNIST. At 1.0 noise probability (pointer~\rpoint{1} in Fig.~\ref{fig:fmnist_qcnn}(a)), the validation accuracy for the Fashion MNIST fluctuates between 70-80\%, while it stabilizes around 75\% with noise levels of 0.1 and 0.2 (pointer~\rpoint{2}). While for other noise levels the QCNN shows performance degradation with increasing noise, at 0.8 and 0.9 noise levels we can observe a highly variable validation accuracy (pointer~\rpoint{3}), suggesting potential robustness with longer training, though risking overfitting. These fluctuations likely stem from dataset complexity and network architecture suitability. 

\paragraph{QCNN robustness against Phase Flip Noise}

As shown in Fig.~\ref{fig:mnist_qcnn}(b) for MNIST and Fig.~\ref{fig:fmnist_qcnn}(b) for Fashion MNIST, the QCNN's response to phase flip varies significantly. For the MNIST, the QCNN with high probabilities ($>$0.6, except 0.9) and at 0.1 probability  (pointer~\rpoint{1} in Fig.~\ref{fig:mnist_qcnn}(b)), the model demonstrates high robustness, maintaining or even exceeding noise-free performance levels. However, for noise probabilities between 0.2 and 0.5 (pointer~\rpoint{2}), as well as at 0.9, the model's accuracy progressively decreases with increasing noise levels. In contrast, the model's performance on the Fashion MNIST shows a highly irregular pattern across different noise levels, with no clear trend of improvement or degradation. 

These inconsistent trends indicate that the QCNN's ability to learn and adapt to noise patterns is highly contingent on specific noise conditions and complexity of the learning task. This behavior likely stems from the interaction between phase flip noise (equivalent to Z gates) and the QCNN's circuit architecture, which incorporates parameterized rotations, entangling operations, and Control-Z gates (depicted in Fig.~\ref{fig:hqnn}) . This complex response highlights the intricate interplay between quantum noise, task complexity, and the QCNN's learning dynamics, underscoring the importance of carefully designing quantum circuits to achieve robust performance in QML tasks.

\paragraph{QCNN robustness against Depolarization Noise}

The performance of QCNNs under depolarization noise is presented in Fig.~\ref{fig:mnist_qcnn}(c) for MNIST and Fig.~\ref{fig:fmnist_qcnn}(c) for Fashion MNIST. Similar to the trends observed in QuanNN, the QCNN exhibits performance degradation with increasing noise levels for both datasets.

For MNIST, at noise levels 0.1 and 0.2 (pointer~\rpoint{1} in Fig.~\ref{fig:mnist_qcnn}(c)), the model demonstrates learning capability, achieving validation accuracies of 90\% and 60\% respectively. However, the QCNN for Fashion MNIST shows more severe degradation, with the model failing to learn at most noise levels. Only at 0.1 and 0.3 noise levels (pointer~\rpoint{1} in Fig.~\ref{fig:fmnist_qcnn}(c)), it shows a slight improvement, more than 50\% validation accuracy. This discrepancy can be attributed to the different complexity of Fashion MNIST classification compared to MNIST, suggesting that higher task complexity amplifies the detrimental impact of depolarization noise. The observed performance degradation aligns with the impact observed in the QuanNN under depolarization noise. As explained in the QuanNN depolarization analysis, this type of noise can lead to rapid loss of quantum coherence, which is essential for maintaining the amount of superposition and entanglement necessary for quantum computation. 

\paragraph{QCNN robustness against Phase Damping Noise}

The performance of QCNNs under both ideal conditions and phase damping noise, presented in Figures~\ref{fig:mnist_qcnn}(d) and~\ref{fig:fmnist_qcnn}(d), reveal non-uniform QCNN performance under phase damping noise for MNIST and Fashion MNIST, respectively. Unlike the QuanNN, the QCNN exhibits varying performance across different noise levels for both datasets. We can identify three distinct performance groups for the model trained on MNIST.

\begin{itemize}[leftmargin=*]
    \item High noise levels (0.7-1.0, pointer~\rpoint{1}): the QCNN fails to learn effectively.
    \item Low noise levels (0.1-0.6, except 0.4, pointer~\rpoint{2} ): the QCNN's accuracy is comparable with the noise-free model.
    \item Noise level 0.4 (pointer~\rpoint{3}): the QCNN outperforms the noise-free model, demonstrating efficient adaptation and resilience to this specific noise intensity.
\end{itemize}

For the Fashion MNIST, similar to the phase flip noise results, the QCNN's performance across different noise levels of phase damping shows no discernible pattern, with highly fluctuating results. These results indicate that the QCNN's response to phase damping is highly dependent on the specific noise level and the nature of the task, highlighting the intricate relationship between quantum noise effects, circuit architecture, and the model's learning dynamics in complex classification tasks.

\paragraph{QCNN robustness against Amplitude Damping Noise}
The performance of QCNNs under amplitude damping noise is presented in Fig.~\ref{fig:mnist_qcnn}(e) for MNIST and Fig.~\ref{fig:fmnist_qcnn}(e) for Fashion MNIST. The results show that the QCNN's performance under amplitude damping noise follows similar patterns observed in the QuanNN, with the model exhibiting various learning behavior at low noise levels for both datasets. 

For the MNIST, the QCNN with noise probabilities between 0.2-0.4 achieves approximately 50-55\% validation accuracy (pointer~\rpoint{1} in Fig.~\ref{fig:mnist_qcnn}(e)). Beyond these levels, the model fails to learn (pointer~\rpoint{2}). For the Fashion MNIST, the QCNN's performance with 0.1 noise level aligns with the noise-free model (pointer~\rpoint{1} in Fig.~\ref{fig:fmnist_qcnn}(e)). However, for amplitude damping noise with probability $\geq$ 0.2 (pointer~\rpoint{2}), the model's performance drastically fluctuates, ranging between 10\% to 30\% validation accuracy, except for a sudden peak at 60\% followed by a drop to 30\% accuracy for the QCNN with 0.5 noise level (pointer~\rpoint{3}). We can also observe that the model tends to learn for 0.4 noise level, reaching 45\% accuracy (pointer~\rpoint{4}). These fluctuating patterns are attributed to the nature of the amplitude damping noise, which directly affects qubit energy levels, challenging quantum algorithms to maintain coherence and perform reliable computations at higher noise levels. Moreover, the QCNN's pooling layers introduce in-circuit measurements, which might further interact with noise and contribute to this behavior. This interplay between noise, circuit architecture, and dataset complexity underscores the intricate dynamics at play in noisy quantum neural networks.

\subsection{Summary of Key Observations and Guidelines}

Our analysis demonstrates how various HQNN architectures respond to specific noise channels and intensities, laying the groundwork for developing more resilient models. We identify critical noise types that significantly impact the performance and determine acceptable mitigation levels for certain architectures. Our results show that the QuanNN exhibits greater robustness than the QCNN across all examined noise types and task demands. Specifically:
\begin{itemize}[leftmargin=*]
    \item The QuanNN demonstrates strong robustness to phase-related noise across all levels, indicating its suitability for devices prone to such errors. 
    \item For readout errors simulated by bit flip, QuanNN shows remarkable resilience at 100\% noise probability, suggesting its robustness against non-deterministic noise and indicating potential need for error-mitigating techniques. 
\end{itemize}

In contrast, the QCNN shows a more nuanced performance across different noise models, showing robustness at specific noise levels for various error types:
\begin{itemize}[leftmargin=*]
    \item Robust performance against bit flip noise at 1.0 intensity.
    \item Robust for phase flip at 0.1, 0.7, 0.8, and 1.0 noise levels.
    \item Consistent performance for phase damping at 0.2 and 0.4 noise levels. 
\end{itemize}

This varied performance implies that QCNN architecture selection should be carefully tailored to the specific quantum device's noise profile.

\section{Conclusion}
\label{sec:conclusion}

This paper presents a comprehensive analysis of HQNNs within the framework of NISQ devices, focusing on their robustness against various quantum noise in image classification tasks. Our evaluation of the QCNN and QuanNN under ideal and noisy conditions revealed significant performance variations with the introduction of quantum noise, particularly across different task complexities, as demonstrated by our results on the MNIST and Fashion MNIST datasets.

Our analysis reveals a complex relationship between noise levels and model performance, highlighting the importance of architecture selection based on specific quantum noise characteristics. While this study advances the understanding of noise effects in QML, future research should explore the interplay between quantum noise and model performance for developing noise-resilient architectures and extend this analysis across a broader range of tasks and datasets.

\section*{Acknowledgements}
This work was supported in parts by the NYUAD Center for Quantum and Topological Systems (CQTS), funded by Tamkeen under the NYUAD Research Institute grant CG008, and the NYUAD Center for Cyber Security (CCS), funded by Tamkeen under the NYUAD Research Institute Award G1104.

\bibliographystyle{ieeetr}
\bibliography{main}
\end{document}